\documentclass[aps,prl,twocolumn,reprint,superscriptaddress,nobibnotes, floatfix]{revtex4-2}
\usepackage{graphicx}
\usepackage{epstopdf}
\usepackage{amssymb, amsmath}
\usepackage[usenames]{color}
\usepackage{bm}
\usepackage{ulem}
\usepackage{multirow}
\usepackage{threeparttable}
\usepackage{booktabs}
\usepackage{dcolumn}
\usepackage{float}
\usepackage[colorlinks=true, linkcolor=blue,anchorcolor=blue, citecolor=blue,urlcolor=blue]{hyperref}

\usepackage[left]{lineno} % 里面的选项代表双栏

\begin{document}
	
\title{Ferroelectric switching of quantum anomalous Hall effects in MnBi$_2$Te$_4$ films}

\author{Jiaheng Li}
\affiliation{Beijing National Laboratory for Condensed Matter Physics and Institute of Physics, Chinese Academy of Sciences, Beijing 100190, China}

\author{Quansheng Wu}
\email{quansheng.wu@iphy.ac.cn}
\affiliation{Beijing National Laboratory for Condensed Matter Physics and Institute of Physics, Chinese Academy of Sciences, Beijing 100190, China}
\affiliation{University of Chinese academy of sciences, Beijing 100049, China}

\author{Hongming Weng}
\email{hmweng@iphy.ac.cn}
\affiliation{Beijing National Laboratory for Condensed Matter Physics and Institute of Physics, Chinese Academy of Sciences, Beijing 100190, China}
\affiliation{University of Chinese academy of sciences, Beijing 100049, China}
\affiliation{Songshan Lake Materials Laboratory, Dongguan, Guangdong 523808, China}

\date{\today}

\begin{abstract}
The integration of ferroelectric and topological materials offers a promising avenue for advancing the development of quantum material devices. In this work, we explore the strong coupling between topological states and ferroelectricity in the heterostructure formed by interfacing MnBi$_2$Te$_4$ (MBT) thin films and monolayer In$_2$Te$_3$. Our first-principles calculations demonstrate that the polarization direction in In$_2$Te$_3$ can strongly alter electronic band structures in the MBT/In$_2$Te$_3$ heterostructure, and even induces a topological phase transition between quantum anomalous Hall ($C=1$) and trivial ($C=0$) insulating states, originating from the change of band order induced by the switch of out-of-plane polarization. Our work highlights the promising potential of ferroelectric-topological heterostructures in aiding the development of reconfigurable quantum devices, and creating new possibilities for progress in advanced microelectronic and spintronic systems.
\end{abstract}

\maketitle

%\linenumbers % 开始编号

Quantum anomalous Hall (QAH) insulators \cite{haldane1988model, chang2023quantum} are characterized by an insulating bulk gap and nondissipative chiral edge channels. The experimental observation of quantized Hall plateau was first realized in magnetically doped topological insulator (TI) (Bi,Sb)$_2$Te$_3$ thin films in the low temperature of tens of millikelvin \cite{chang2013experimental}. Following this groundbreaking discovery, several devices made from magnetically doped TIs have been demonstrated to exhibit the QAH effect \cite{chang2015high, chang2015zero, mogi2015magnetic, ou2018enhancing}, where the consequent magnetic disorder hinders the robust observation of the QAH effect at high temperatures.

\begin{figure}[htbp]
	\includegraphics[width=0.8\linewidth]{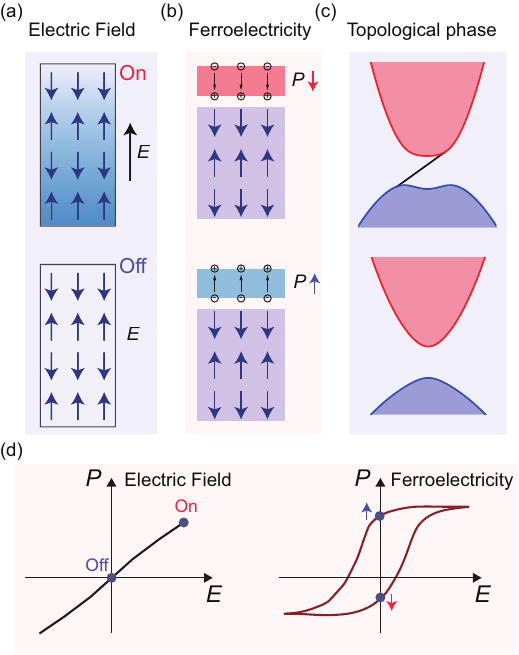}
	\caption{Schematic diagrams of topological phase transition induced by electric field and ferroelectric polarization. (\textbf{a}) The ``on'' and ``off'' states induced by external electric field. (\textbf{b}) The ``up'' and ``down'' states with opposite ferroelectric polarization.  (\textbf{c}) Schematic diagrams of bulk and edge states of quantum anomalous Hall insulating and trivial insulators.  (\textbf{d}) The response of polarization versus external electric field, where binary states are manipulated by electric field and ferroelectric polarization.}
	\label{fig_TPT}
\end{figure}

The MnBi$_2$Te$_4$ (MBT) single crystal, exhibiting the A-type antiferromagnetic (AFM) order up to 25 K \cite{yan2019crystal, lee2019spin}, has been recently theoretically predicted \cite{li2019intrinsic, zhang2019topological, otrokov2019unique} and experimentally fabricated \cite{lee2013crystal, gong2019experimental, otrokov2019prediction, cui2019transport, li2020antiferromagnetic}, which inherits the van der Waals nature and overcomes the disadvantage of magnetic disorder caused by doping. The unique A-type AFM order and its interplay with topology provide MBT as an ideal platform to explore exotic topologically related physical phenomena, such as the long-sought antiferromagnetic topological insulator\cite{mong2010antiferromagnetic, li2019intrinsic, zhang2019topological, gong2019experimental, otrokov2019prediction}, layer-dependent topological properties with QAH and axion insulating behaviors \cite{li2019intrinsic, otrokov2019unique, deng2020quantum, liu2020robust, li2025stacking, lian2025antiferromagnetic}, axion electrodynamics \cite{sekine2021axion, zhao2021routes, qiu2025observation}, layer Hall effect \cite{gao2021layer, chen2022layer}, high-Chern-number QAH insulators \cite{ge2020high, bosnar2023high, li2025high}, nonlinear Hall effect \cite{gao2023quantum, wang2023quantum}, and other intriguing physical phenomena \cite{li2019magnetically, ceccardi2023anomalous, xu2024second, yang2025septuple}.

Meanwhile, topological phase transitions (TPTs) can be triggered by hydrostatic pressure \cite{bahramy2012emergence}, alloying \cite{xu2012observation}, and strain engineering \cite{huang2016tensile, mutch2019evidence}. Specifically, the idea of an electric-field-induced TPT (Fig. \ref{fig_TPT} (a)) is crucial for developing topological field-effect transistors \cite{qian2014quantum, chong2023electric, zhang2021heterobilayer, marrazzo2022twist}, where the ``on'' state allows for ballistic charge or spin transport along edge states, and the ``off'' state suppresses this channel. However, this approach is less effective for storage devices. Ferroelectric materials (Fig. \ref{fig_TPT} (b)), with their persistent ``up'' and ``down'' polarization states, are better suited for non-volatile memory applications. The integration of ferroelectric control into topological materials (Fig. \ref{fig_TPT} (c-d)) not only opens new possibilities for the emergence of topological memory devices, but also holds the potential to transform the field of data storage and information processing.

% The concept of topological order in condensed matter physics has fundamentally reshaped our understanding of various phases of matter, Topological physics has been a crucial area of condensed matter physics that explores the unique properties of materials based on their topological characteristics, leading to phenomena like topological insulators with surface states that are robust against disorder. Various experimental setups, including hydrostatic pressure, alloying, magnetic fields, strain engineering, and electric fields, can effectively induce topological phase transitions in materials.

In the work, we demonstrate that heterostructure formed by MnBi$_2$Te$_4$ (MBT) 4-septuple (SL) film and In$_2$Te$_3$ yields the strong coupling between ferroelectricity and topological states. Through first-principles calculations, we theoretically predict that switching the polarization direction in In$_2$Te$_3$ can manipulate the electronic structure of the MBT/In$_2$Te$_3$ heterostructure, inducing transitions between quantum anomalous Hall ($C=1$) and trivial insulating ($C=0$) states. Layer-resolved anomalous Hall conductivity (AHC) calculations reveal that local AHC is predominantly  concentrated within the MBT films, and that the switch of out-of-plane polarization notably alters the real-space distribution of AHC. These findings underscore the potential for designing innovative quantum devices by harnessing the strong coupling between ferroelectric states and topology in MBT/In$_2$Te$_3$ heterostructures.

\textit{Method}--First-principles calculations are performed using the projector augmented wave method in the framework of density functional theory (DFT) implemented in the Vienna $ab initio$ simulation package (VASP) \cite{kresse1996efficient}. The ionic environment is simulated by the Perdew-Burke-Ernzerhof (PBE) exchange correlation potential \cite{perdew1996generalized} with the generalized gradient approximation. The energy cutoff of plane wave basis is fixed at 400 eV, and all the structures are fully relaxed with a force criterion of 0.01 eV / \r{A}. The localized $d$-orbitals on Mn atoms are tackled by the DFT + $U$ method with $U = 4 \mathrm{eV}$, which is consistent with the previous paper \cite{li2019intrinsic, li2019magnetically}. The van der Waals (vdW) correction is included by adopting the DFT-D3 method \cite{grimme2010consistent}. Topological invariants and surface state calculations are calculated using the \texttt{WannierTools} package \cite{wu2018wanniertools}, based on tight-binding Hamiltonians constructed from maximally localized Wannier functions \cite{mostofi2014updated}.

\begin{figure}[htbp]
	\includegraphics[width=1.0\linewidth]{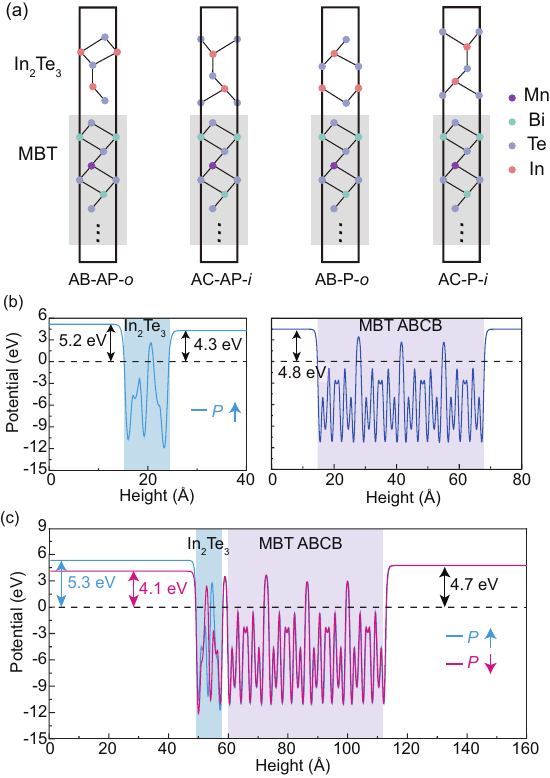}
	\caption{(\textbf{a}) Typical MnBi$_2$Te$_4$ (MBT)-ABCB/In$_2$Te$_3$ heterostructures with distinct stacking patterns, such as AB-AP-$o$, AC-AP-$i$, AB-P-$o$ and AC-P-$i$, where the relative stacking between MBT and In$_2$Te$_3$ is denoted by ``AB'' and ``AC'', the relative stacking direction is denoted by parallel (``P'') and antiparallel (``AP''), and the polarization of In$_2$Te$_3$ is denoted by $i$ (inwards) and $o$ (outwards). (\textbf{b}) Local potential in monolayer In$_2$Te$_3$ with polarization aligning outwards and MBT 4-SL ABCB, with reference to the Fermi level.(\textbf{c}) Local potential in the MBT-ABCB/In$_2$Te$_3$ heterostructures with $i$ and $o$ polarization, with reference to the Fermi level.}
	\label{fig_str_wf}
\end{figure}

As a typical vdW two-dimensional material, bulk MBT crystalline in the $R\bar{3}m$ structure with the ABC stacking order with the building block in the Te-Bi-Te-Mn-Te-Bi-Te sequence \cite{lee2013crystal, gong2019experimental, otrokov2019prediction} and can be further exfoliated into thin films with arbitrary layer number from its bulk form and exhibit various stacking patterns through interlayer sliding. Another monolayer, In$_2$Te$_3$ (with Te-In-Te-In-Te quintuple atom layers) \cite{serebryanaya1992crystal}, has been theoretically predicted to be a ferroelectric insulator with moderate energy barrier and dynamical stability \cite{ding2017prediction}. Both films possess an in-plane triangular lattice structure, and the close match of their lattice constants (within $1\%$), with $a=4.366$ \r{A} for In$_2$Te$_3$ and $a=4.36$ \r{A} for MBT, facilitating the experimental integration of these two films in the heterostructure.

\begin{figure*}[htbp]
	\includegraphics[width=1.0\linewidth]{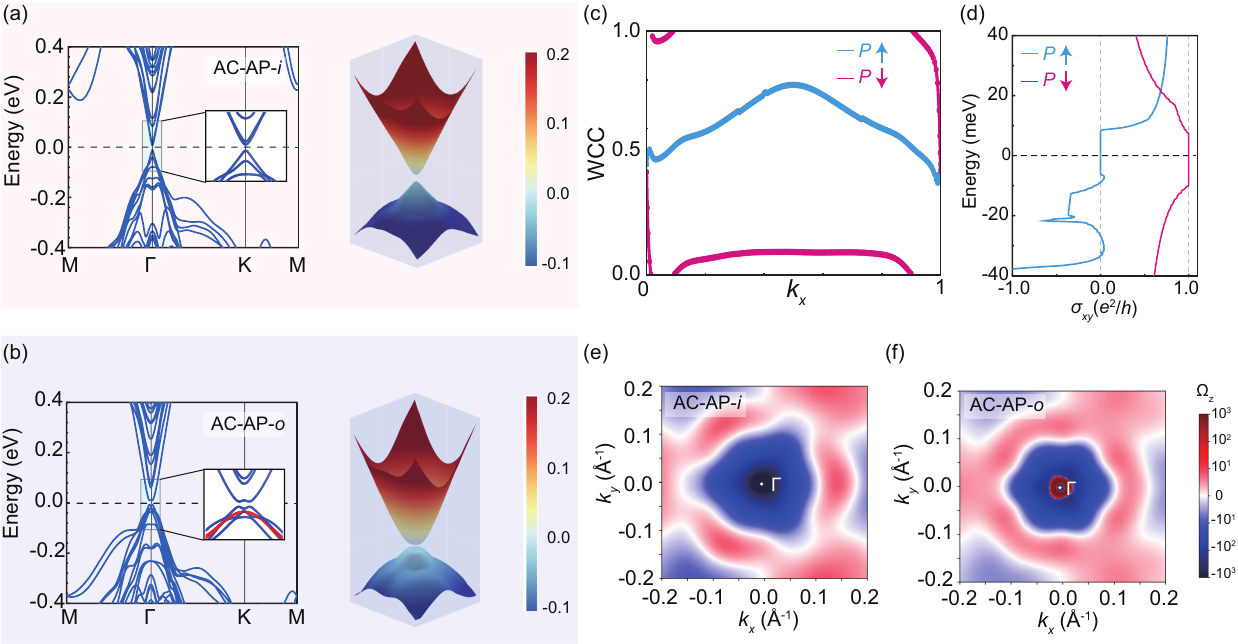}
	\caption{Electronic and topological properties of MBT/In$_2$Te$_3$ heterostructure with polarization inwards and outwards, denoted by AC-AP-$i$ and AC-AP-$o$. (\textbf{a}) and (\textbf{b}) Electronic band structures along high-symmetry lines and  energy contours of valence and conduction bands in the AC-AP-$o$ and AC-AP-$i$ heterostructure, with the projected band structures shown in the inset, where red and blue colors denote the contribution of In$_2$Te$_3$ and MBT ABCB 4-SL slab. (\textbf{c}) and (\textbf{d}) The evolution of Wannier charge centers (WCCs) and anomalous Hall conductivity in the AC-AP-$o$ and AC-AP-$i$ case. (\textbf{e}) and (\textbf{f}) Momentum-resolved Berry curvature ($\Omega_z$) within the zoomed Brillouin zone around $\Gamma$.}
	\label{fig_band}
\end{figure*}

MBT 4SL with the ABCB stacking pattern can be realized by sliding the topmost SL in the exfoliated 4-SL ABCA film. The lateral shift makes MBT ABCB become QAH insulators with $C=1$ \cite{li2025stacking} and the non-equivalence between the top and bottom surfaces. Various stacking configurations are systematically explored, and it is found that high-symmetry positions are energetically favored, where the most stable configuration follows the ABC stacking order with relative in-plane shift of (0, 0), (1/3, -1/3), (2/3, -2/3). The typical crystal structure of MBT/In$_2$Te$_3$ heterostructure is shown in Fig. \ref{fig_str_wf} (a). To distinguish various stacking patterns of MBT/In$_2$Te$_3$ heterostructure, we denote the polarization of In$_2$Te$_3$ by $i$ (inwards) and $o$ (outwards), the relative stacking position between MBT/In$_2$Te$_3$ by ``AB'' and ``AC'', and the relative stacking direction by parallel (``P'') and antiparallel (``AP''). Therefore, the combination of these symbols provides a comprehensive framework for iterating through all potential stacking configurations, such as AB-AP-$o$ and AC-P-$i$ in Fig. \ref{fig_str_wf} (a).

In our subsequent analysis, we concentrate on the antiparallel (AP) stacking direction, among which the AC-AP-$i$ configuration is the most energetically favorable, and other stacking configurations can be found in the supplementary material \cite{SM}. The exfoliation energy in these heterostructures is defined as $E_{\mathrm{exfoliation}} = E_{\mathrm{MBT/In_2Te_3}} - E_{\mathrm{MBT}} - E_{\mathrm{In_2Te_3}}$, and it is consistently less than 3 meV /\r{A}$^2$ for all configurations. This value is in well accordance with the typical exfoliation energies observed for 2D van der Waals materials \cite{bjorkmen2012van, mounet2018two}, indicating that the interface between the MBT/In$_2$Te$_3$ heterostructure are primarily held together by van der Waals interaction. Furthermore, the negligible difference in exfoliation energy between AB- and AC-stacked heterostructures suggests that achieving in-plane shifts is feasible in experimental settings \cite{wu2021sliding}. However, there arises a significant difference in heterostructures with the polarization in In$_2$Te$_3$ orienting inwards and outwards, whose numerical value is approximately 2 meV /\r{A}$^2$. The physical origin of this energy difference in two stable ferroelectric states can be attributed to asymmetric crystal structure in the heterostructure and the difference of real-space charge distribution on both sides of In$_2$Te$_3$ induced by polarization switching.

As a typical member of the ferroelectric insulator family, In$_2$Te$_3$ monolayer exhibits finite out-of-plane polarization, resulting in the electrostatic potential energy difference of approximately 0.9 eV on both surfaces, as illustrated in Fig. \ref{fig_str_wf} (b). In contrast, the MBT 4-SL ABCB slab, despite lacking spatial inversion ($\mathcal{I}$) and horizontal mirror plane ($\mathcal{M}_z$) symmetries, shows a constant work function across its top and bottom surfaces, indicating negligible intrinsic out-of-plane polarization, compared to typical 2D ferroelectric materials, such as In$_2$Se$_3$ and CuInP$_2$Se$_6$ \cite{ding2017prediction, maisonneuve1997ferrielectric, qi2021widespread}.

When MBT 4-SL and In$_2$Te$_3$ monolayer further form the MBT/In$_2$Te$_3$ heterostructure in Fig. \ref{fig_str_wf} (c), the polar structure of In$_2$Te$_3$ gives rise to distinct vacuum levels on the surface of In$_2$Te$_3$, and electrostatic potential energy difference on both surfaces of the MBT/In$_2$Te$_3$ heterostructure in Fig. \ref{fig_str_wf} (c) remains $\phi\approxeq 0.6$ eV, independent of polarization direction in In$_2$Te$_3$. The work function of MBT 4SL is compatible with the surfaces of In$_2$Te$_3$, and the band gap of In$_2$Te$_3$ ($\Delta_g\approxeq 1.0$ eV) is larger than the electrostatic energy difference, demonstrating MBT/In$_2$Te$_3$ heterostructure with a type-I band alignment, characterized by a distinct straddling gap without any energetic overlap between the valence and conduction band of the heterostructure constituents \cite{ihn2009semiconductor}.

The computed electronic band structures of MBT/In$_2$Te$_3$ heterostructure with polarization inwards and outwards is shown in Fig. \ref{fig_band} (a-b). Both band dispersions in MBT/In$_2$Te$_3$ heterostructures with the AC-AP-$o$ and AC-AP-$i$ stacking patterns exhibit the insulating behavior and the valence and conduction band maximum are located around $\Gamma$, inheriting the typical characteristics in the electronic states of MBT ABCB films.

The projected band structure in the inset of Fig. \ref{fig_band} (a) and (b) highlights the contribution of atomic orbitals from MBT and In$_2$Te$_3$ to electronic states around the Fermi level, where red and blue colors denote the contribution of In$_2$Te$_3$ and MBT ABCB 4-SL slab. In the AC-AP-$i$ structure, the highest valence and conduction bands mainly originate from the MBT 4-SL, which resemble massive Dirac cones typical of surface states on bulk AFM MBT, suggesting that the AC-AP-$i$ structure retains a similar topological character to MBT AFM bulk. The similar massive Dirac cones can also be observed in the AC-AP-$o$ configuration. Conversely, in the AC-AP-$o$ configuration, the In$_2$Te$_3$ monolayer significantly contributes to the valence bands near the Fermi level, creating a unique "M"-shaped dispersion. This significant difference in electronic states suggests band inversion at $\Gamma$, indicating a topological phase transition induced by polarization switching.

To verify whether ferroelectricity polarization-induced topological phase transition occurs in MBT/In$_2$Te$_3$ heterostructure, we explicitly compute the Chern numbers of the AC-AP-$o$ and AC-AP-$i$ structures by using the Wilson loop method which traces the evolution of Wannier charge centers in all the occupied states. The obtained Chern numbers in Fig. \ref{fig_band} (c) are $C=0$ and $C=1$ for the outwards and inwards polarization. The anomalous Hall conductivity (AHC) $\sigma_{xy}$ in Fig. \ref{fig_band} (d) also validate that switching out-of-plane polarization gives rise to distinct topological insulating states with a full gap exceeding 10 meV, suggesting the strong coupling between ferroelectricity and band topology.

The calculated momentum-resolved Berry curvature ($\Omega_z(\mathbf{k})$) shown in Fig. \ref{fig_band} (e) and (f) in both structures exhibits three-fold rotation symmetry and is primarily concentrated around $\Gamma$, which inherits a characteristic feature of MBT films. The most significant differences in Berry curvature are located in the interior Brillouin zone around $\Gamma$, which is the consequence of band inversion at $\Gamma$ caused by polarization switching, and can be tracked back to distinct orbital hybridization at the interface between the In$_2$Te$_3$ monolayer and the MBT slabs shown in inset of Fig. \ref{fig_band} (a) and (b). Thus, the MBT/In$_2$Te$_3$ heterostructures behave as a ferroelectric QAH insulators, where the ``up'' and ``down'' states can be effectively manipulated through external electric field, similar to its family material In$_2$Se$_3$.

\begin{figure}[htbp]
	\includegraphics[width=1.0\linewidth]{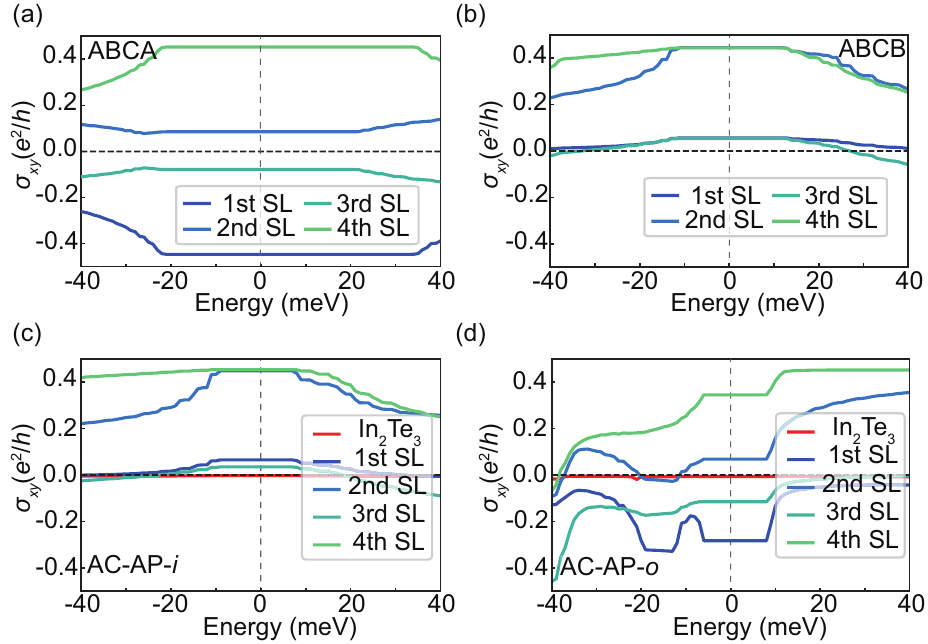}
	\caption{Layer-resolved anomalous Hall conductivity (AHC) as a function of chemical potential in MBT 4SL thin films and MBT/In$_2$Te$_3$ heterostructures, where each MBT SL and In$_2$Te$_3$ are denoted by different colors. (\textbf{a} and \textbf{b}) Layer-resolved AHC in MBT ABCA- and ABCB-stacking thin film. (\textbf{c} and \textbf{d}) Layer-resolved AHC in MBT/In$_2$Te$_3$ heterostructure with polarization inwards and outwards.}
	\label{fig_LAHC}
\end{figure}

In order to clarify the role of In$_2$Te$_3$ monolayer in this ferroelectric QAH insulator and understand how different layers in the MBT/In$_2$Te$_3$ heterostructure contribute to the overall AHC, the layer-resolved AHC can be computed by decomposing the Chern-Simons contribution to AHC in the real space \cite{bianco2011mapping, marrazzo2017locality, varnava2018surface}, which can be defined as

\begin{equation}
	\begin{aligned}
	\sigma_{xy}({\alpha}) = -\frac{e^2}{h} \int_{\mathrm{BZ}}d\mathbf{k} \mathrm{Tr}[\mathcal{P}_\mathbf{k} x \mathcal{Q}_{\mathbf{k}} y],
	\end{aligned}
\end{equation}
where $\mathcal{P}_\mathbf{k}=\sum_j \theta(\mu - \epsilon_{j, \mathbf{k}}) |\psi_{j\mathbf{k}}\rangle \langle\psi_{j\mathbf{k}}|$ is the ground-state projector into the occupied states at $\mathbf{k}$, and $\mathcal{Q}_\mathbf{k} = \mathcal{I}-\mathcal{P}_\mathbf{k}$. The Fermi-Dirac distribution is introduced to tackle with the insulating and metallic case in the same framework \cite{marrazzo2017locality}.

Let us reconsider the original MBT 4-SL film without interfacing with In$_2$Te$_3$ first. The existence of the combination operator of spatial inversion ($\mathcal{I}$) and time-reversal ($\mathcal{T}$) symmetry constrains MBT 4-SL films to possess $C=0$, and the largest contributions to AHC are observed in the top and bottom SLs in Fig. \ref{fig_LAHC} (a), originating from unique topological properties in bulk MBT materials \cite{li2019intrinsic, zhang2019topological, otrokov2019unique}. The lateral shift of the topmost layer from ABCA to ABCB-stacking in MBT 4-SL film weakens the interlayer coupling between the topmost SL and the remaining SLs, inducing a topological phase transition from an axion insulator with $C=0$ to a QAH insulator with $C=1$ in the ABCB stacking \cite{li2025stacking}. Conversely, the lateral shift in the topmost SL results in the largest AHC observed in the 2nd and 4th SLs, with the topmost SL contributing relatively little, shown in Fig. \ref{fig_LAHC} (b). This stacking dependent topology highlights the significance of interlayer coupling on electronic and topological properties in MBT family materials.

When monolayer In$_2$Te$_3$ with opposite polarization interfaces with MBT ABCB 4-SL slabs, the real space distribution of AHC shown in Fig. \ref{fig_LAHC} (c, d) is primarily concentrated in the MBT region, with minimal contribution from In$_2$Te$_3$ layer. Furthermore, the layer-dependent AHC in the AC-AP-$i$ case is similar to that observed in the MBT-ABCB films. Contrarily, when the polarization in In$_2$Te$_3$ switches to the outward direction, the layers exhibiting the most pronounced contribution to AHC become the top and bottom surfaces, and layer-resolved AHC from neighboring SLs with the opposite signal, similar to the MBT ABCA 4-SL slab. Overall. the unusual behavior observed in layer-resolved AHC can provide valuable insights into polarization induced topological phase transition in the MBT/In$_2$Te$_3$ heterostructure.

% Combining the results from projected band structures and layer-resolved AHC, the reversal of ferroelectric polarization in In$_2$Te$_3$ causes different orbital hybridization at the interface between monolayer In$_2$Te$_3$ and MBT slabs, which is universal in heterostructures formed by 2D materials. Thus, the layer-resolved AHC clearly underscore the critical role of ferroelectric polarization in the In$_2$Te$_3$ layer in shaping the electronic and topological properties of magnetic topological materials.

\textit{Discussion} -- In this study, we predict the heterostructure composed of MBT 4-SL ABCB slab and monolayer In$_2$Te$_3$ is a ferroelectric quantum anomalous Hall (QAH) insulator. Our first-principles calculations demonstrate that polarization switching in In$_2$Te$_3$ can induce a topological phase transition, toggling between a QAH insulating state ($C=1$) and a trivial insulating state ($C=0$). The projected band structures and layer-resolved anomalous Hall conductivity reveal that ferroelectric polarization strongly alters orbital hybridization at the interface, and induces band inversion at $\Gamma$. While only MBT/In$_2$Te$_3$ heterostructures with MBT-4SL are suitable for ferroelectric-switching topology \cite{SM},  similar ferroelectric switching of topology can be expected to exist in heterostructures formed by other ferroelectric materials (such as In$_2$Te$_3$, CuInP$_2$Se$_6$) and MBT family materials (like MnSb$_2$Te$_4$, MnBi$_4$Te$_7$). Our works demonstrate the strong interplay between ferroelectric polarization and topology, underscores the potential for manipulating topological properties in MBT/In$_2$Te$_3$ heterostructure through external electric fields, and suggests the promising potential for integrating ferroelectric and topological materials to create non-volatile topological memory devices.

\begin{acknowledgments}
\textit{Acknowledgments.---}This work was supported by the National Key R\&D Program of China (Grant No. 2022YFA1403800, 2023YFA1607401), the National Natural Science Foundation of China (Grant No.12274436, 11925408, 11921004), the Science Center of the National Natural Science Foundation of China (Grant No. 12188101) and  H.W. acknowledge support from the Informatization Plan of the Chinese Academy of Sciences (CASWX2021SF-0102) and the New Cornerstone Science Foundation through the XPLORER PRIZE. J. L. acknowledge support from China National Postdoctoral Program for Innovation Talents (Grant No. BX20220334).
\end{acknowledgments}

\section{Contributions}
J.L., Q.W., and H.W. conceived the project. J. L. performed the first-principles calculations, analyzed raw data and wrote the initial manuscript with the help of Q.W., and H.W. All authors were engaged in discussing the main results and editing the final manuscript.

\section{Data availability}
Data are available from the corresponding authors upon reasonable request.

\section{Ethics declarations}
\subsection{Competing interests}
The authors declare no competing interests.

\end{document}